\documentclass[epj,nopacs]{svjour}
\newif\ifpdf\ifx\pdfoutput\undefined\pdffalse\else\pdfoutput=1\pdftrue\fi
\newcommand{\pdfgraphics}{\ifpdf\DeclareGraphicsExtensions{.pdf, .jpg}\else\DeclareGraphicsExtensions{.eps, .ps, .jpg}\fi}
\usepackage{graphicx}

\begin{document}

\pdfgraphics
\title{\boldmath Constraining the Unitarity Triangle with $B\to K^{*}\gamma$ and $B\to\rho\gamma$}
\author{Stefan W. Bosch
}                     
\institute{Institute for High-Energy Phenomenology, Newman Laboratory for Elementary-Particle Physics,\\ Cornell University, Ithaca, NY 14853, U.S.A. \hfill {\tt CLNS 03/1847}} %
\date{Received: date / Revised version: date}
%
\abstract{
We discuss the exclusive radiative decays $B\to K^{*}\gamma$ and $B \to\rho\gamma$ in QCD factorization within the Standard Model. The analysis is based on the heavy-quark limit of QCD. Our results for these decays are complete to next-to-leading order in QCD and to leading order in the heavy-quark limit. Phenomenological implications for branching ratios and isospin breaking effects are discussed. Special emphasis is placed on constraining the CKM unitarity triangle from these observables.
%
} 
\titlerunning{Constraining the Unitarity Triangle with $B\to K^{*}\gamma$ and $B\to\rho\gamma$}
\maketitle

\section{Introduction}
\label{intro}
The rare radiative $B$ decays belong to the most valuable probes of the 
quark flavour sector (see \cite{review} for a recent review). The inclusive 
$b\to s\gamma$ mode, showing good agreement of the theoretical 
next-to-leading-logarithmic (NLL) QCD prediction and experimental measurements,
puts stringent bounds on physics beyond the standard model. CP-averaged branching ratios of exclusive radiative channels are measured to be $B(B^{0} \to K^{*0}\gamma) = (4.18 \pm 0.23) \cdot 10^{-5}$ and $B(B^{+} \to K^{*+}\gamma) = (4.14 \pm 0.33) \cdot 10^{-5}$ \cite{BKgamexp}, and bounded with 90\% confidence 
level as $B(B^{0} \to \omega^{0}\gamma) < 1.0 \cdot 10^{-6}$, 
$B(B^{0} \to \rho^{0}\gamma) < 1.2 \cdot 10^{-6}$ and 
$B(B^{+} \to \rho^{+}\gamma) < 2.1 \cdot 10^{-6}$ \cite{Brhoexp}.

Whereas the inclusive decay can be treated perturbatively, bound state effects are essential for the exclusive modes and have to be described by some nonperturbative quantities like hadronic form factors and light-cone distribution amplitudes (LCDAs). However, in the heavy-quark limit $m_{b}\gg\Lambda_{\mathrm{QCD}}$
a systematic treatment of exclusive $B$ decays is possible within 
QCD \cite{BBNS}: Perturbatively calculable contributions to the matrix elements can be factorized from nonperturbative form factors and universal light-cone distribution amplitudes. We use the QCD factorization technique for the 
exclusive radiative decays $B\to K^{*}\gamma$ and $B\to\rho\gamma$ as in 
\cite{BFS,BB,AP}. 
The ratio of the $B\to\rho\gamma$ and $B\to K^{*}\gamma$ branching fractions is, at leading order in $\alpha_{s}$, directly proportional to the side $R_{t}$ 
in the standard unitarity triangle (UT), where
\begin{equation}
R_t\equiv\sqrt{(1-\bar\rho)^2+\bar\eta^2}=
\frac{1}{\lambda}\left|\frac{V_{td}}{V_{ts}}\right|
\end{equation}
Having the complete NLL result for the decay amplitudes in $B\to V\gamma$ 
at hand, we can calculate $\alpha_{s}$ corrections to their relation 
with $R_t$ and evaluate the implications in the ($\bar\rho,\bar\eta)$ plane 
\cite{AliLunghi,UTBVgam}.

\section{$B\to V\gamma$ at NLO in QCD}
\label{sec:BVgamNLO}
The effective weak Hamiltonian for $b\to s/d\gamma$ transitions is
\begin{equation}\label{heff}
  {\cal H}_{eff}=\frac{G_F}{\sqrt{2}}\sum_{p=u,c}\lambda_p^{(s/d)}
\bigg( \sum_{i=1}^2 C_i Q^p_i +\sum_{j=3}^8 C_j Q_j\bigg)
\end{equation}
where $\lambda_p^{(s/d)}=V^*_{ps/d}V_{pb}$. The relevant operators are the current-current operators $Q^p_{1,2}$, the QCD-penguin operators $Q_{3\ldots 6}$, and the electro- and chromomagnetic penguin operators $Q_{7,8}$. To evaluate the hadronic matrix elements of these operators we employ the heavy-quark limit 
$m_{b}\gg\Lambda_{\mathrm{QCD}}$ to get the factorization formula \cite{BFS,BB}
\begin{eqnarray}\label{fform}
  \lefteqn{\langle V\gamma(\epsilon)|Q_i|\bar B\rangle =} \\
  &&\hspace*{-4ex} =\Big[ F^{B\to V} T^I_{i} + \!\int^1_0 \!\!d\xi\, dv \, T^{II}_i(\xi,v) \Phi_B(\xi) \Phi_V(v)\Big] \!\cdot\epsilon \nonumber
\end{eqnarray}
where $\epsilon$ is the photon polarization 4-vector. Here $F^{B\to V}$ is a $B\to V$ transition form factor, and $\Phi_B$, $\Phi_V$ are leading-twist light-cone distribution amplitudes of the $B$ meson and the vector meson $V$, respectively. These quantities are universal, nonperturbative objects. They describe the long-distance dynamics of the matrix elements, which is factorized from the perturbative, short-distance interactions expressed in the hard-scattering kernels $T^I_{i}$ and $T^{II}_i$. To leading order in QCD and leading power in the heavy-quark limit, $Q_7$ gives the only contribution to the $B\to V\gamma$ amplitude. At ${\cal O}(\alpha_s)$ the operators $Q_{1\ldots 6}$ and $Q_8$ start contributing and the factorization formula becomes nontrivial.

\begin{figure}
\center{\resizebox{0.48\textwidth}{!}{%
  \includegraphics{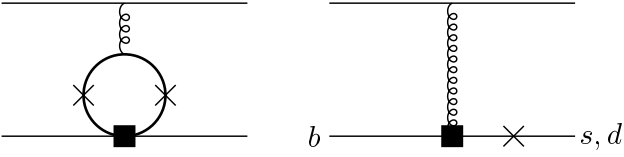}
}}
\caption[1]{\label{fig:qit2} \it ${\cal O}(\alpha_s)$ contribution at 
leading power to the hard-scattering kernels $T^{II}_i$ from four-quark 
operators $Q_i$ (left) and from $Q_8$. The crosses indicate the places 
where the emitted photon can be attached.}
\end{figure}

The relevant diagrams for the NLO hard-vertex corrections $T_i^I$ were computed in \cite{GHWBCMU} to get the virtual corrections to the matrix elements for the inclusive $b\to s\gamma$ mode at next-to-leading order. 
For the exclusive modes the same corrections play the role of
the perturbative type I hard-scattering kernels. The non-vanishing 
contributions to $T^{II}_i$ where the spectator participates in the 
hard scattering are shown in Fig.~\ref{fig:qit2}. 
We can express both the type I and type II contributions to the matrix elements $\langle Q_i\rangle$ in terms of the matrix element $\langle Q_7\rangle$, an explicit factor $\alpha_s$, and hard-scattering functions $G_i$ and $H_i$ which are given explicitely in \cite{BB,thesis}.

Weak annihilation contributions are suppressed by one power of $\Lambda_{\mathrm{QCD}}/m_{b}$ but they are nevertheless calculable in QCD factorization because in the heavy-quark limit the colour-transparency argument applies to the emitted, highly energetic vector meson. Including them we become sensitive to the charge of the decaying 
$B$ meson and thus to isospin breaking effects.

\section{Results}
\label{sec:results}
\begin{figure}[t]
\center{\resizebox{0.48\textwidth}{!}{%
\includegraphics{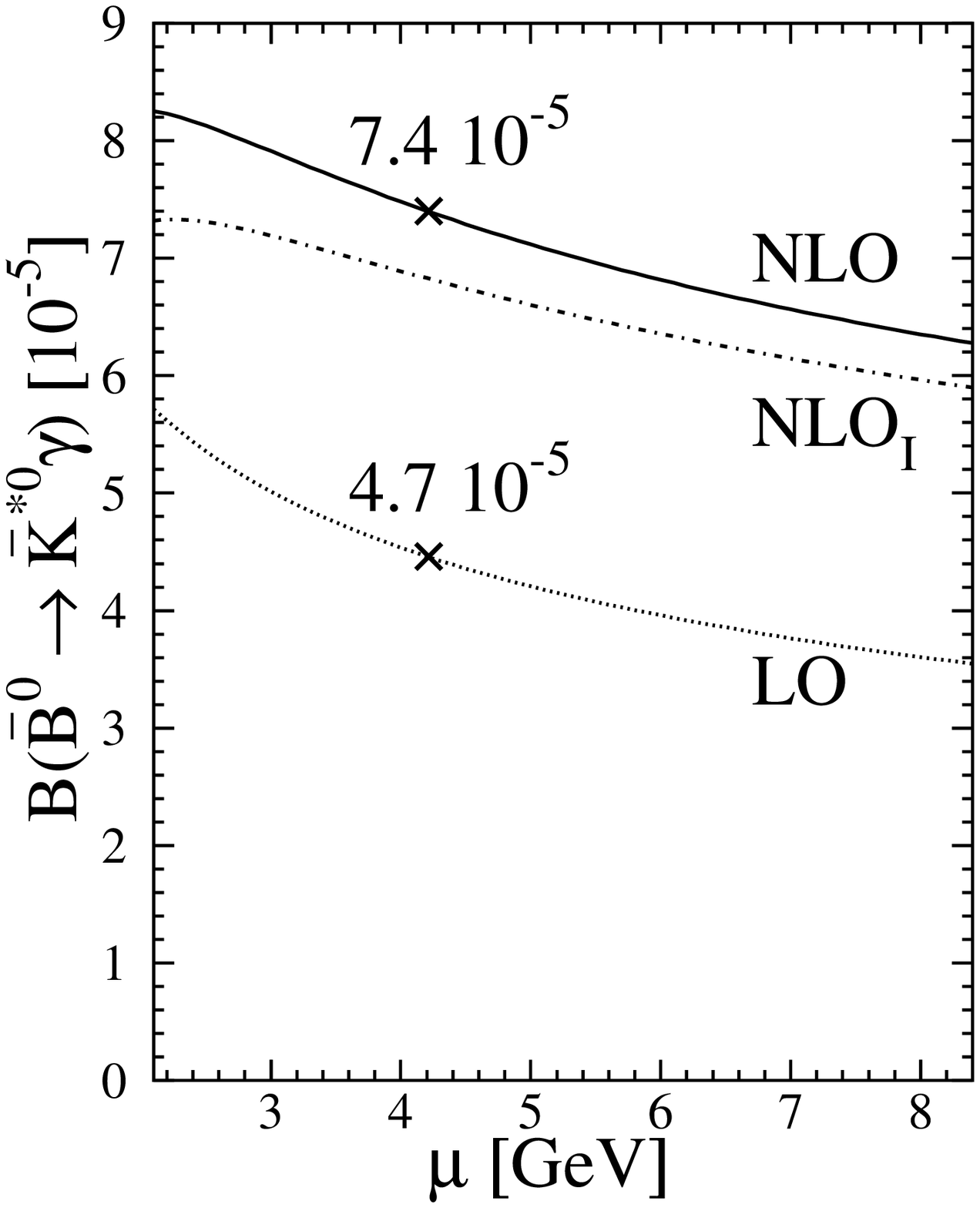}\hspace{0.2cm}\includegraphics{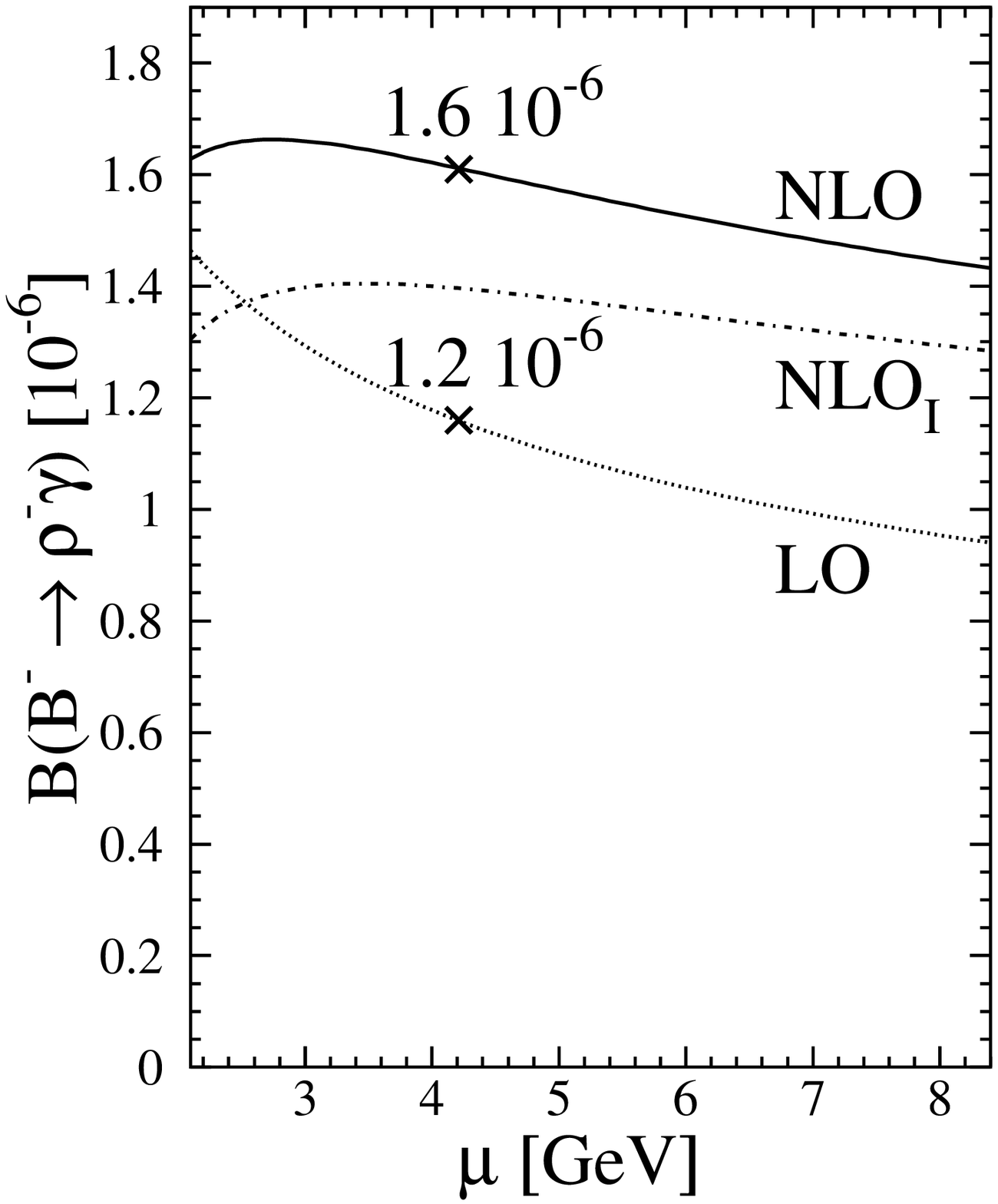}
}}
\caption[1]{\it Dependence of the branching fractions $B(\bar{B}^0 \to \bar{K}^{*0} \gamma)$ and $B(B^- \to \rho^- \gamma)$ on the renormalization scale $\mu$. The dotted line shows the LO, the dash-dotted line the NLO result including type-I corrections only and the solid line shows the complete NLO result.}
\label{fig:bkrhomu}
\end{figure}

The total $\bar B\to V\gamma$ amplitude then can be written as
\begin{equation}
\label{amp}
A(\bar B\to V\gamma)=\frac{G_F}{\sqrt{2}}\left[\lambda_u a^u_7 +
\lambda_c a^c_7\right]\langle V\gamma|Q_7|\bar B\rangle
\end{equation}
where the factorization coefficients $a_7^p(V\gamma)$ consist of the Wilson coefficient $C_7$, the contributions from the type I and type II hard-scattering and annihilation corrections. One finds a sizeable enhancement of the leading order value, dominated by the $T^I$-type correction. The net enhancement of $a_7$ at NLO leads to a corresponding enhancement of the branching ratios, for fixed value of the form factor. This is illustrated in Fig. \ref{fig:bkrhomu},
where we show the residual scale dependence for $B(\bar{B}\to \bar{K}^{*0}\gamma)$ and $B(B^-\to\rho^-\gamma)$ at leading and next-to-leading order. 
Our central values for the $B\to K^{*}\gamma$ branching ratios are 
higher than the quoted experimental measurements. The dominant uncertainty in the theoretical values comes from the $B\to V\gamma$ form factors. We used the light-cone sum rule (LCSR) results $F_{K^{*}}=0.38\pm 0.06$ and 
$F_{\rho}=0.29\pm 0.04$ from \cite{BB98}. 
A recent preliminary lattice QCD determination, 
$F_{K^{*}}=0.25\pm 0.05\pm 0.02$ \cite{SPQCDR}, would give a better 
agreement with the experimental central values.

The charge averaged isospin breaking ratio can be defined as
\begin{equation}
\Delta (V\gamma)=\frac{\Gamma (B^{0}\to V^{0}\gamma) - v \Gamma(B^{\pm}\to V^{\pm}\gamma)}{\Gamma (B^{0}\to V^{0}\gamma) + v \Gamma(B^{\pm}\to V^{\pm}\gamma)}
\end{equation}
with $v=1$ for $V=K^{*}$ and $v=1/2$ for $V=\rho$.
This ratio has a reduced sensitivity to the nonperturbative form factors. 
As already discussed, in our approximations, isospin breaking is generated by weak annihilation contributions. Kagan and Neubert found a large effect from the penguin operator $Q_{6}$ on the isospin asymmetry $\Delta (K^{*}\gamma)$ \cite{KN}. The prediction $\Delta(K^*\gamma)=(3.9^{+3.1}_{-1.9})\%$ 
is in agreement with the experimental value 
$\Delta(K^*\gamma)_{exp}=(4.5\pm 4.9)\%$. For $B\to\rho\gamma$ we find a 
strong dependence of the isospin asymmetry on the angle $\gamma$ of the 
unitarity triangle. The $\gamma$ dependence is in particular pronounced 
for the zero crossing of $\Delta (\rho\gamma)$ around $\gamma=60^{\circ}$, 
the value favoured by the standard UT fits.

\section{Implications for the UT Analysis}
\label{sec:UT}

We can use observables in the $B\to V\gamma$ decay modes to get information 
on parameters in the ($\bar\rho,\bar\eta$) unitarity triangle plane. The 
cleanest observable is the ratio of the neutral $B\to\rho\gamma$ and 
$B\to K^{*}\gamma$ branching ratios. (We choose here the neutral modes because 
annihilation effects could be sizeable in $B^{\pm}\to\rho^{\pm}\gamma$.
However, those effect can be estimated and the charged mode
can also be used for a similar analysis.) 

\begin{figure}[t]
\center{\resizebox{0.48\textwidth}{!}{%
\includegraphics{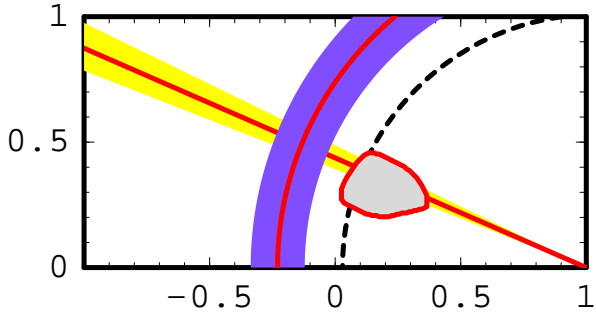}
}}
\caption[1]{\it Impact of the experimental upper bound on $R_{00}$ in the 
($\bar\rho,\bar\eta$) plane. The width of the dark band reflects the variation 
of $\xi$. The intersection with the 
light-shaded $\sin (2\beta )$ band defines the apex of the unitarity triangle 
and the length $R_{t}$.} 
\label{fig:R00}
\end{figure}

\begin{figure}[t]
\center{\resizebox{0.48\textwidth}{!}{%
\includegraphics{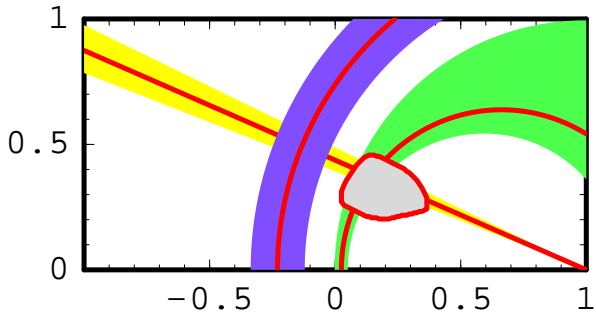}
}}
\caption[1]{\it Same as Fig.~\ref{fig:R00} including the implication of a 
measurement of $\Delta (\rho\gamma)^{\mbox{exp}}=0$ (curved band on the right).
The width of the  
band reflects the theoretical uncertainties from varying the hadronic 
parameter $\lambda_{B}$ and the renormalization scale $\mu$. 
(The effect of isospin breaking in the form factors is neglected here.)}
\label{fig:R00iso}
\end{figure}

We define
\begin{eqnarray}
\label{R00def}
R_{00} &\equiv& \frac{B(B^{0}\to\rho^{0}\gamma)}{B(B^{0}\to K^{*0}\gamma)}\\
\label{R00}
  &\approx& \frac{1}{2} \left| \frac{V_{td}}{V_{ts}}\right|^{2} \xi^{-2} 
\left(1+\Delta(\bar\rho,\bar\eta) \right)
\end{eqnarray}
where CP averaged branching fractions are understood.
Here $\Delta(\bar\rho,\bar\eta)$  is a small perturbative correction 
\cite{UTBVgam} and $\xi=F_{K^{*}}/F_{\rho}$, the ratio of the form factors, is essentially the only source of theoretical uncertainty. We use the LCSR 
estimate $\xi =1.33\pm 0.13$ \cite{BB98}. A preliminary lattice value is $\xi =1.1\pm 0.1$ \cite{SPQCDR}. Experimentally, so far only an upper limit on $R_{00}$ exists. Because the $B\to\omega\gamma$ branching ratio is up to tiny corrections the same as the one for $B^{0}\to\rho^{0}\gamma$ \cite{UTBVgam} and its experimental limit is tighter, we use it to get $R_{00}^{\mbox{exp}}<0.024$. 
If we use in addition the experimental measurement $\sin (2\beta)=0.734\pm 0.054$ \cite{sin2b} we can construct the overlap of the $R_{00}$ and $\sin(2\beta)$ bands and extract the length $R_{t}$ as shown in Fig.~\ref{fig:R00}.
The dashed curve in Fig.~\ref{fig:R00} was obtained setting 
$\xi=1$ in (\ref{R00}). This can be viewed as the lowest possible value.
If $R_{00}$ were measured at its current experimental bound
$R_{00}=0.024$, the dashed line would correspond to a conservative
lower bound on $R_t$.
The value $R_{t}<1.24$ from 
$R_{00}$ is already becoming comparable with the constraints from 
$\Delta M_{B_{d}}$ and $\Delta M_{B_{s}}$. It is possible that the 
experimental measurement of $R_{00}$ 
may actually be achieved before the measurement of $\Delta M_{B_{s}}$. 
Once a measurement of both the charged and neutral $B\to\rho\gamma$ modes 
is available, one can also use $\Delta(\rho\gamma)$ to constrain the 
unitarity triangle. For illustration purposes we plot in Fig.~\ref{fig:R00iso} 
in addition to the $R_{00}$ and $\sin (2\beta)$ bands the implication of an 
assumed measurement of $\Delta (\rho\gamma)^{\mbox{exp}}=0$, which 
would correspond to the Standard Model prediction
for a CKM angle $\gamma =60^{\circ}$.

\section{Conclusions and Outlook}
\label{sec:conc}

We have discussed a systematic and model-independent NLL framework for the 
rare radiative decays $B\to V\gamma$ based on the heavy-quark 
limit $m_b\gg\Lambda_{\mathrm{QCD}}$. As observables of primary
interest we considered the 
branching fractions, the ratio $R_{00}$ of the neutral 
$B\to\rho\gamma$ and $B\to K^{*}\gamma$ branching fractions, and the isospin 
breaking ratios $\Delta (K^{*}\gamma)$ and $\Delta(\rho\gamma)$. Our main 
focus was on the implications of measurements of these quantities on the 
($\bar\rho,\bar\eta$) plane of the CKM unitarity triangle. The theoretically 
cleanest quantities are $R_{00}$ and $\Delta(\rho\gamma)$ which, however, are 
not yet measured experimentally. We hope that they will become accessible
in the near future with new data from the $B$ factories.

{\em Acknowledgements:} These results were obtained in collaboration with Gerhard Buchalla to whom I'm very grateful. I would like to thank Sebastien Descotes-Genon, Thor\-sten Feldmann, Bj\"orn Lange, and Enrico Lunghi for interesting discussions during this conference. This research was supported by the National Science Foundation under Grant PHY-0098631.

\end{document}